\documentclass[twocolumn,showpacs,superscriptaddress,pra]{revtex4-1}
\usepackage{graphicx}
\usepackage{times}
\usepackage{textcomp}
\usepackage{amsmath,amssymb}
\usepackage{tikz}
\usepackage{color}
\usetikzlibrary{positioning,arrows}
\usetikzlibrary{decorations.pathmorphing}
\usetikzlibrary{decorations.markings}
\definecolor{myblue}{RGB}{56,94,141}

\begin{document}
\newcommand{\newc}{\newcommand}
\newc{\beq}{\begin{equation}}
\newc{\eeq}{\end{equation}}
\newc{\kt}{\rangle}
\newc{\br}{\langle}
\newc{\beqa}{\begin{eqnarray}}
\newc{\eeqa}{\end{eqnarray}}
\newc{\pr}{\prime}
\newc{\longra}{\longrightarrow}
\newc{\ot}{\otimes}
\newc{\rarrow}{\rightarrow}
\newc{\h}{\hat}
\newc{\bom}{\boldmath}
\newc{\btd}{\bigtriangledown}
\newc{\al}{\alpha}
\newc{\be}{\beta}
\newc{\ld}{\lambda}
\newc{\sg}{\sigma}
\newc{\p}{\psi}
\newc{\eps}{\epsilon}
\newc{\om}{\omega}
\newc{\mb}{\mbox}
\newc{\tm}{\times}
\newc{\hu}{\hat{u}}
\newc{\hv}{\hat{v}}
\newc{\hk}{\hat{K}}
\newc{\ra}{\rightarrow}
\newc{\non}{\nonumber}
\newc{\ul}{\underline}
\newc{\hs}{\hspace}
\newc{\longla}{\longleftarrow}
\newc{\ts}{\textstyle}
\newc{\f}{\frac}
\newc{\df}{\dfrac}
\newc{\ovl}{\overline}
\newc{\bc}{\begin{center}}
\newc{\ec}{\end{center}}
\newc{\dg}{\dagger}
\newc{\prh}{\mbox{PR}_H}
\newc{\prq}{\mbox{PR}_q}
\newc{\tr}{\mbox{tr}}
\newc{\pd}{\partial}
\newc{\qv}{\vec{q}}
\newc{\pv}{\vec{p}}
\newc{\dqv}{\delta\vec{q}}
\newc{\dpv}{\delta\vec{p}}
\newc{\mbq}{\mathbf{q}}
\newc{\mbqp}{\mathbf{q'}}
\newc{\mbpp}{\mathbf{p'}}
\newc{\mbp}{\mathbf{p}}
\newc{\mbn}{\mathbf{\nabla}}
\newc{\dmbq}{\delta \mbq}
\newc{\dmbp}{\delta \mbp}
\newc{\T}{\mathsf{T}}
\newc{\J}{\mathsf{J}}
\newc{\sfL}{\mathsf{L}}
\newc{\C}{\mathsf{C}}
\newc{\B}{\mathsf{M}}
\newc{\V}{\mathsf{V}}
\title{Protocol using kicked Ising dynamics for generating states with maximal multipartite entanglement}
\author{Sunil K. Mishra}
\email{sunilkm.app@iitbhu.ac.in}
\affiliation{ Department of Physics, Indian Institute of Technology, Banaras Hindu University, Varanasi - 221005, India}
\author{Arul Lakshminarayan}
\email{arul@physics.iitm.ac.in}
\affiliation{Department of Physics, Indian Institute of Technology Madras, Chennai -600036, India}
\author{V. Subrahmanyam}
\email{vmani@iit.ac.in}
\affiliation{Department of Physics, Indian Institute of Technology Kanpur UP - 208016, India}
\begin{abstract}
We present a solvable model of iterating cluster state
protocols that lead to entanglement production, 
between contiguous blocks, of 1 ebit per iteration. This continues till the blocks are maximally entangled at which stage an
 unravelling begins at the same rate till the blocks are unentangled. The model is a variant of the transverse field Ising model
 and can be implemented with CNOT and single qubit gates. The inter qubit entanglement as measured by the concurrence is shown to be zero
 for periodic chain realizations while for open boundaries there are very specific instances at which these can develop. Thus we introduce a class
 of simply produced  states with very large multipartite entanglement content of potential use in measurement based quantum computing.
 \end{abstract}
\maketitle

Entanglement has been recently studied very extensively as it impacts both foundational as well as applied aspects of quantum 
theory (for a review see \cite{RMPH4}). Along with studies aimed at understanding the nature of entanglement, its presence in condensed matter
 systems, especially spin chains, has been used as a good platform to study various aspects of many-body systems including quantum phase
 transitions \cite{bose,subram1,latorre,osborne,amico,diptiman}. 
 One such widely studied system is the Ising model
 which exhibits a quantum phase
 transition when a magnetic field is applied in a direction transverse to the interaction \cite{sachdev}.
 A variant of the transverse-field Ising model, when the field is applied impulsively kicked via a Dirac delta comb, has been discussed 
in the literature \cite{prosen, arul1,sunil1}. Even in the presence of time dependence the transverse Ising model remains integrable via the
 Jordan-Wigner transformation \cite{jordan} as the resulting fermions are free. 
These models can lead to very highly entangled states with multipartite entanglement. 

Mutlipartite entanglement, whose understanding is still incomplete, has been investigated in many recent works \cite{Brunner2012,Arnaud2013}.
 The generalized cluster states, or graph states, with highly persistent distributed entanglement, has been proposed as a model of quantum 
computation, distinct from the circuit model, the so-called one-way quantum computer \cite{Raussendorf2001,Nielsen2005,Nielsen2006}. Cluster states 
have also been experimentally obtained, for example see \cite{Lu2007} and have been applied for constructing quantum error correcting codes, which for
 example has been realized experimentally recently \cite{Bell2014,Hein2004}.
Here, we construct spin states with high multipartite entanglement using the kicked Ising model dynamics, by iterating the
quantum-circuit protocol with the cluster states \cite{Briegel2001}. While graph states are constructed with qubits being the vertices of a graph 
and the edges representing Ising interactions, we may look upon the present
 study as what happens in the simplest case when the action of the gates is {\it iterated} on the resultant graph states.
 This iteration leads to nontrivial multipartite entangled states, as the single qubit vertex operations do not commute with the interaction-based two-qubit ones,  as will be seen below. 
For a study of closely related Hamiltonians and more motivation see \cite{sunil1}, especially from the quenching point of view.
As the model considered here is closely allied to the the dynamics of quantum critical transverse-field Ising model, it
can  provide  motivation and interest for the structure of the resultant many-body states.
It may be noted in passing 
that that cluster states do not naturally occur as ground states of Hamiltonians \cite{Nielsen2006},  but the iterated cluster states, that we 
introduce here,  are natural to obtain via time evolution.

 Consider the Hamiltonian for a spin chain of $L$ sites, 
\begin{equation}
  {\cal H}(t)=\sum_{j=1}^{L-1} \sigma_j^x \sigma_{j+1}^x +\sum_{k=-\infty}^{\infty}\delta(k-\frac{t}{\tau}) \sum_{j=1}^L \sigma_j^z 
\label{fixed_ham}
\end{equation}
when the period of the pulses is $\tau=\pi/4$.  We will study the dynamics of an open spin chain here, the periodic chain
can similarly be discussed.
The propagator or unitary Floquet quantum map  (in the sense of for example \cite{balazsberryvorostabor}) connecting states just before any two 
consecutive kicks is given by   
\begin{eqnarray}
 U=\exp\left(-i\frac{\pi}{4}\sum_{j=1}^L\sigma_j^x\sigma_{j+1}^x\right) \exp\left(-i\frac{\pi}{4}\sum_{j=1}^L\sigma_j^z\right)
\label{kick_map}
\end{eqnarray}
The operator $\exp(i\frac{\pi}{4}\sigma_1^x\sigma_2^x)$ is the Cartan form of a CNOT gate \cite{marcin,vidal}. Thus we can represent and implement quite simply the evolution operator as a quantum circuit
comprising of CNOT gates and one-qubit gates.
  At time $t=n\tau$, where $\tau=\frac{\pi}{4}$ and $n$ is total number of kicks, the time evolved state is $U^n|0\cdots0\rangle$, where $|0\rangle$ is the eigenstate of local $\sigma^z$ operator.
 The non-triviality of the time evolution stems from the non-commutativity of the operators involving different Pauli matrices 
in Eq.~(\ref{kick_map}). If the $\sigma_z$ single qubit operators were not present, the resulting state for $n=1$ was first 
studied as a ``cluster state" \cite{Briegel2001}.

\begin{figure}[t]
{\hglue -0.3cm
\includegraphics[width=1.\linewidth]{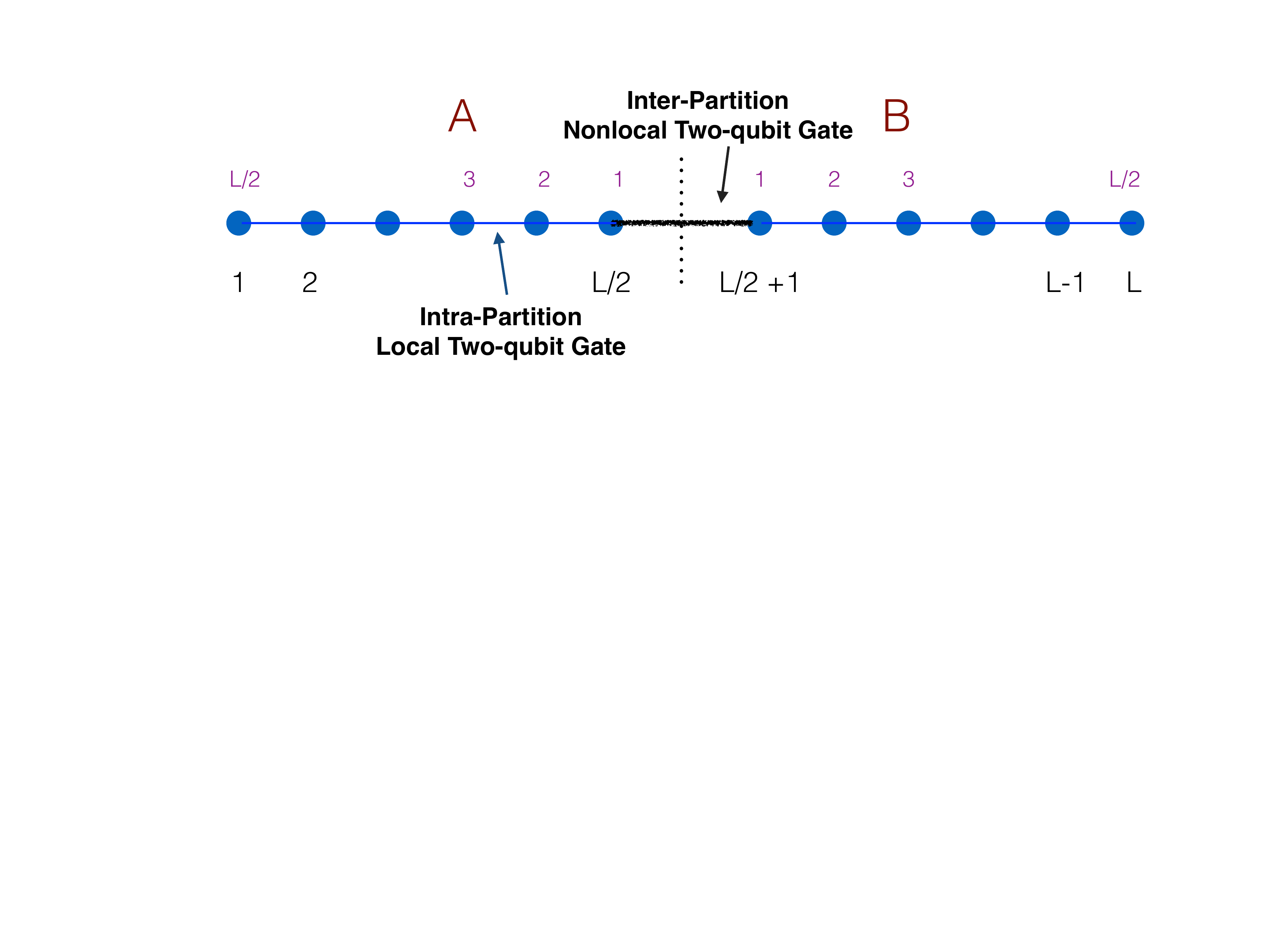}
}
\vskip -3 cm
\caption{The partitioning of the spin chain chain into two blocks  $A$ and $B$,  with relabelling of the sites. }
\label{fig0}
\end{figure}

We divide the spin chain into two blocks $A$ with $M$ spins and $B$ with $N=L-M$ spins.
For simplicity, we consider the case of  $M=L/2$ first.
The blocking of the lattice, along with relabeling of the sites is as shown in Fig.~\ref{fig0} for the open boundary conditions.
Let us introduce the following notation for spin operators of different partitions,  for $ j\le M=L/2$,
\begin{equation}
\vec  A_j\equiv\vec \sigma_{M+1-j},~{\rm and}~
\vec B_j \equiv \vec \sigma_{M+j}.
\label{label_AB}
\end{equation}
Thus $\vec A_1$ and $\vec B_1$ represent the Pauli matrices for spins at locations $M$ and $M+1$ respectively.

The entanglement between the two
 blocks is calculated from the entropy of block $A$ (by tracing out the block $B$ spins). In Fig.~\ref{fig1} the evolution of this entanglement in $|\psi_n\kt$ is shown  for open and periodic boundary 
conditions with $L=20$ spins and equal blocks with $M=10$. In a very remarkable way the entanglement increases with every iteration
 exactly by 1 ebit (entanglement between a maximally-entangled pair of qubits) in the open chain case and by 2 ebits for the  periodic chain. Thus after $L/4$ ($L/2$) iterations, blocks $A$ and $B$ are maximally entangled 
in the periodic (open) chains, with an entanglement of $L/2$ ebits. In the periodic case, the translational symmetry
 ensures that this is the case for {\it every} block of $L/2$ consecutive spins, which makes it a very nontrivial state. Further iterations
 reverse the entanglement and lead to unentangled ones at $L/2$ and $L$ iterations for the case of periodic and open chains respectively.

The unitary operator in Eq.~(\ref{kick_map}) can be written using this notation and labeling as,
\begin{equation}
 U=X_{AB}X_{AA}X_{BB}Z_{A}Z_{B}\equiv U_A U_B V_1,
\label{uxaa}
\end{equation}
 where $X_{AB}=e^{-i\frac{\pi}{4}A_1^xB_1^x}$,  $X_{AA}=e^{-i\frac{\pi}{4}\sum_{j=1}^{L/2-1}A_j^xA_{j+1}^x}$ and
 $Z_A=e^{-i\frac{\pi}{4}\sum_{j=1}^{L/2}A_j^z}$, with similar definitions for $X_{BB}$ and $Z_B$. The operator $U_A\equiv X_{AA} Z_A$ is a multi-spin unitary, 
that uses one-qubit and two-qubit gates, acting on spins of partition $A$ only, the interaction bond between the two partitions is represented by
\begin{equation}
V_1\equiv U_A^{\dagger}U_B^{\dagger}X_{AB}U_AU_B= {\rm e}^{-i \f{\pi}{4}A_1^y B_1^y }.
\end{equation}
In this format the unitary operator's nonlocal parts, as far as the partition $A:B$, is concerned are laid threadbare. 
\begin{figure}[t]
{\hglue -1.1cm
\includegraphics[height=7.5cm,width=1.\linewidth]{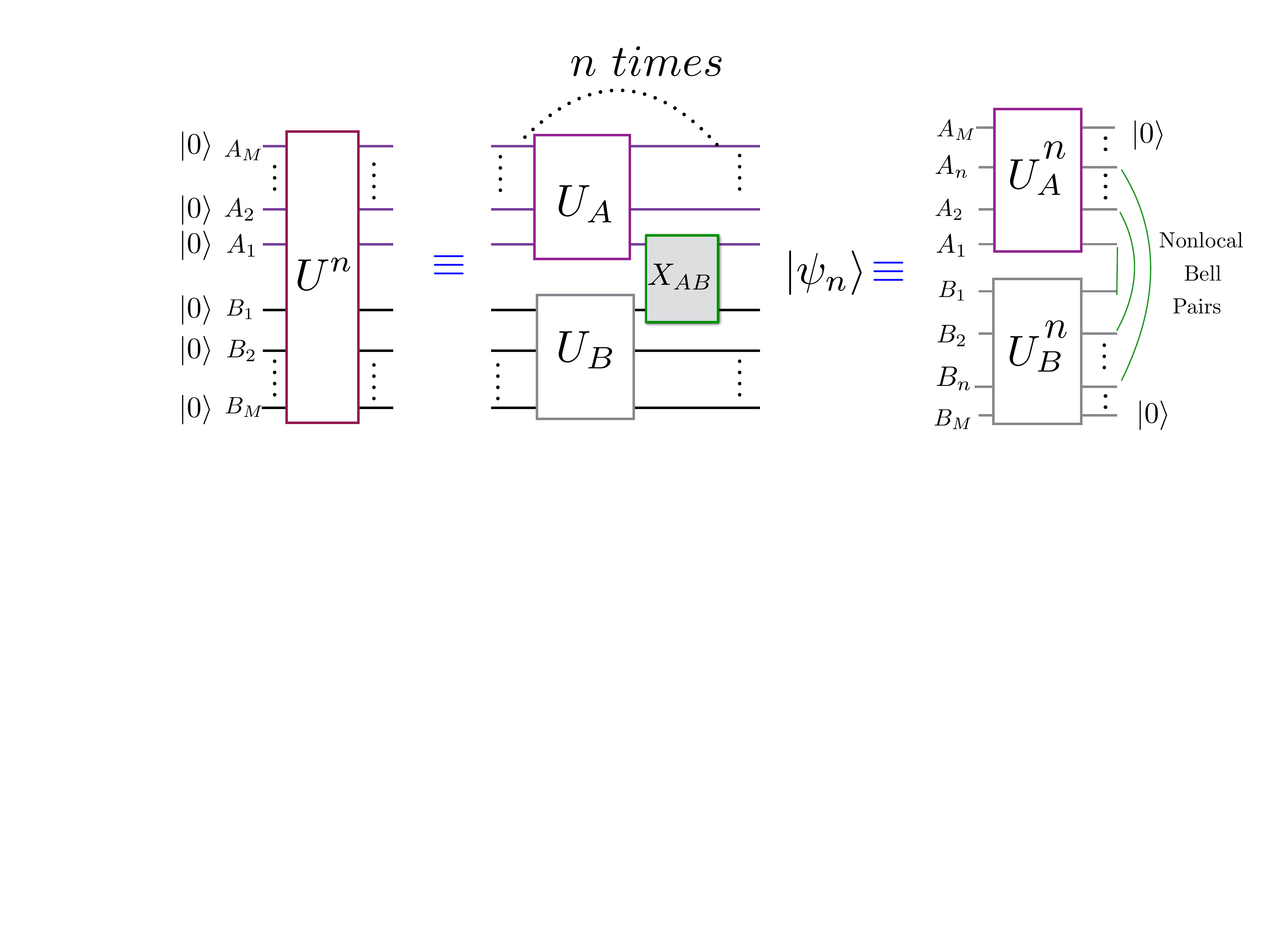}
}
\vskip -4cm
\caption{ Quantum circuit implementing the unitary operation, and the resultant state after n steps. 
The state has exactly $n$ nonlocal Bell pairs, apart from local unitaries that act on individual blocks.}
\label{fig2}
\end{figure}
Now, the powers of the unitary operator can be written as
\beq
U^n = U_A^nU_B^nV_nV_{n-1}V_{n-2}\cdots V_1
\label{vN_eq}
\eeq
where $V_n=U_A^{\dagger}U_B^{\dagger}V_{n-1} U_A U_B= {U_A^{\dagger}}^n{U_B^{\dagger}}^nX_{AB}{U_A}^n{U_B}^n$. 

This can serve as a protocol for generating multipartite nonlocal entanglement using only one nonlocal two-qubit gate $X_{AB}$, acting 
always on the same two qubits. Given that $|\psi_0\kt =\otimes^L |0\kt$, it helps to find the states during the first few steps of the iteration. 
The quantum circuit implementing  this unitary operation in Eq.~(\ref{uxaa}) for $n$ steps on the 
initial unenetangled state $|\psi_0\rangle$ is shown in Fig.~\ref{fig2}, along with the resultant multipartite entangled state 
with $n$ nonlocal (inter-partition) Bell pairs. It is helpful
to also define the state $|\tilde \psi_n \kt = \prod_{i=1}^n V_i |\psi_0\kt$, which is well defined as the $V_i$ are commuting operators as exhibited below.

We will calculate the von Neumann entropy $S_{M}(n)$ of the block with $M$ spins as function of the time steps $n$. 
The state after the first step is straightforward to evaluate, only $V_1$ involves interaction between the partitions.
The state after one time step is, 
$|\psi_1\kt=U |\psi_0\rangle
\equiv U_A U_B |\tilde\psi_1\rangle$, has exactly one ebit of entanglement distributed over all spins. Defining a Bell state between a pair of spins (located at $l$ and $m$) as, $|\Phi_{lm}\rangle \equiv (|00\rangle -i |11\rangle)/\sqrt{2}$, we have,
\begin{equation}
|\tilde \psi_1\rangle =|0..0\rangle_{A_2..A_M}|\Phi_{A_1,B_1}\rangle |0..0\rangle_{B_2..B_M}.
\label{first-step}
\end{equation}
In the state $|\tilde \psi_1\rangle$, the spins $A_1$ and $B_1$ at the interface between the blocks are in a Bell state, which
has one ebit of entanglement. The state $|\psi_1\rangle$  has  additional local unitary transformations $U_A$ and $U_B$ that act on individual blocks,  which do not change the entanglement between the blocks.
However, the two-party entanglement (calculated by the concurrence measure of entanglement \cite{Wootters1998}) between
spins $A_1$ and $B_1$ is zero. This implies that the entanglement in the state $|\psi_1\rangle$ is of multipartite nature.
Thus, we can use this protocol as a reverse-engineering tool to transform a local Bell pair entanglement into a nonlocal multipartite entanglement. Further, a local Bell pair
entanglement can also be transferred to another local Bell pair, by changing the blocking scheme after generating
the state $|\psi_1\rangle$, so that a different pair of spins is at the interface of blocks $A^\prime$ and $B^\prime$, through
$ |\tilde \psi_1^\prime\rangle = U_{A^\prime}^\dagger  U_{B^\prime}^\dagger |\psi_1\rangle.$

Turning to the calculation of $|\psi_2\kt$, the state at the second step, the
operator $V_2$ is required.  The special value  $\tau=\pi/4$ renders a remarkably simple and transparent 
form for this operator (As shown in the Supplementary Material \cite{suppl}):
\beq
V_2=\exp\left(-i \f{\pi}{4} A_{2}^yB_2^y A_1^z B_1^z\right). 
\eeq
In the state $ |\psi_2\kt=U_A^2U_B^2V_2V_1|\psi_0\rangle \equiv U_A^2U_B^2|\tilde \psi_2\rangle$,
 while $V_1$ flips the spins labelled $A_1$ and $B_1$, $V_2$ {\it retains} these while flipping $A_2$ and $B_2$,  thus we have,
\begin{equation}
|\tilde \psi_2\rangle =|0..0\rangle_{A_3..A_M}|\Phi_{A_1,B_1}\Phi_{A_2,B_2}\rangle|0..0\rangle_{B_3..B_M}.
\end{equation}
The reduced density matrix $\rho_A$ has four equal eigenvalues $\lambda=1/4$ and  the von Neumann entropy $S_{L\over2}(n=2)=2$ for the open chain as seen in
 Fig.~\ref{fig1}.  In the periodic boundary  case we get von Neumann entropy of 4 ebits, as every time step increases the
 entropy by two units.
The von Neumann entropy for general $n$ such that $2\le n\le M$ in the open chain can be derived by observing that (see details in the Supplementary Material \cite{suppl})
\beq
V_n = \exp\left(-i \f{\pi}{4} A^y_n B^y_n \prod_{j=1}^{n-1}A^z_{j} B^z_{j}  \right),
\eeq
which may be proved for example by induction. Thus with every iteration the operator $V$ acquires an additional 
two-spins in the interaction. This implies that the state $|\psi_n\rangle\equiv U_A^{n}U_B^{n}|\tilde \psi_n \kt $ is given by,
\begin{equation}
|\tilde \psi_n\rangle =|0..0\rangle_{A_{n+1}..A_M}|\Phi_{A_1,B_1}..\Phi_{A_n,B_n}\rangle|0..0\rangle_{B_{n+1}..B_M},
\label{tilde_psi_n}
\end{equation}
and the entanglement entropy for this state is $n$ ebits. Thus in the open chain case when $n=M=L/2$, the maximum 
entanglement of $L/2$ ebits is achieved. In the closed chain, this time is reduced to $L/4$.

 Beyond the time when the maximum entropy is reached, the expression for $V_n$ is to be modified:
In the open chain case with $M=L/2$,  the operator $V_{M+1}$ is given by (Details in the Supplementary Material \cite{suppl}),
\beq
 V_{M+1}= \exp\left(-i \f{\pi}{4} A^x_{M} B^x_{M} \prod_{j=1}^{M-1}A^z_{j} B^z_{j} \right)
 \label{eq:VLby2p1}
\eeq
and subsequently till $n=L$, the operators with the highest index get decimated such that 
at $V_{M+k}$ ($1 \le k \le M$) the operator string involved is $A^x_{M-k+1}B^x_{M-k}A^z_{M-k}B^z_{M-k}\cdots A^z_{1}B^z_{1}$.
It follows that $V_{L}=\exp(-i \pi A_1^x B_1^x/4)$ and $V_{L+1}=V_1$. Thus the interaction picture operators $V_i$ are periodic, and it is also easy to check that they all commute with each other. 

These can be used along with the observation that $\exp(-i \pi \sigma_1^x \sigma_2^x/4)(|00\kt+i |11\kt)/\sqrt{2}=|00\kt$ to arrive at the state 
after $M+1$ kicks as $|\psi_{M+1}\kt=U_A^{M+1}U_B^{M+1} |\tilde \psi_{M+1}\rangle $, 
\beq
|\tilde \psi_n\rangle =|0\rangle_{A_M} |\Phi_{A_1,B_1}\cdots\Phi_{A_{M-1},B_{M-1}} \rangle |0\rangle_{B_M}
\eeq
which is block-local unitarily equivalent to $|\psi_{M-1}\kt$ and thus the  
entanglement is $M-1$ ebits which is consistent with the open boundary case in Fig.~\ref{fig1}. Thus further time evolution unravels the entanglement at the rate of $1$ ebit  per iteration, and in the periodic boundary case at the rate of $2$ ebits per iteration till there is no entanglement at all.
 \begin{figure}[t]
\includegraphics [angle=0,width=.8\linewidth] {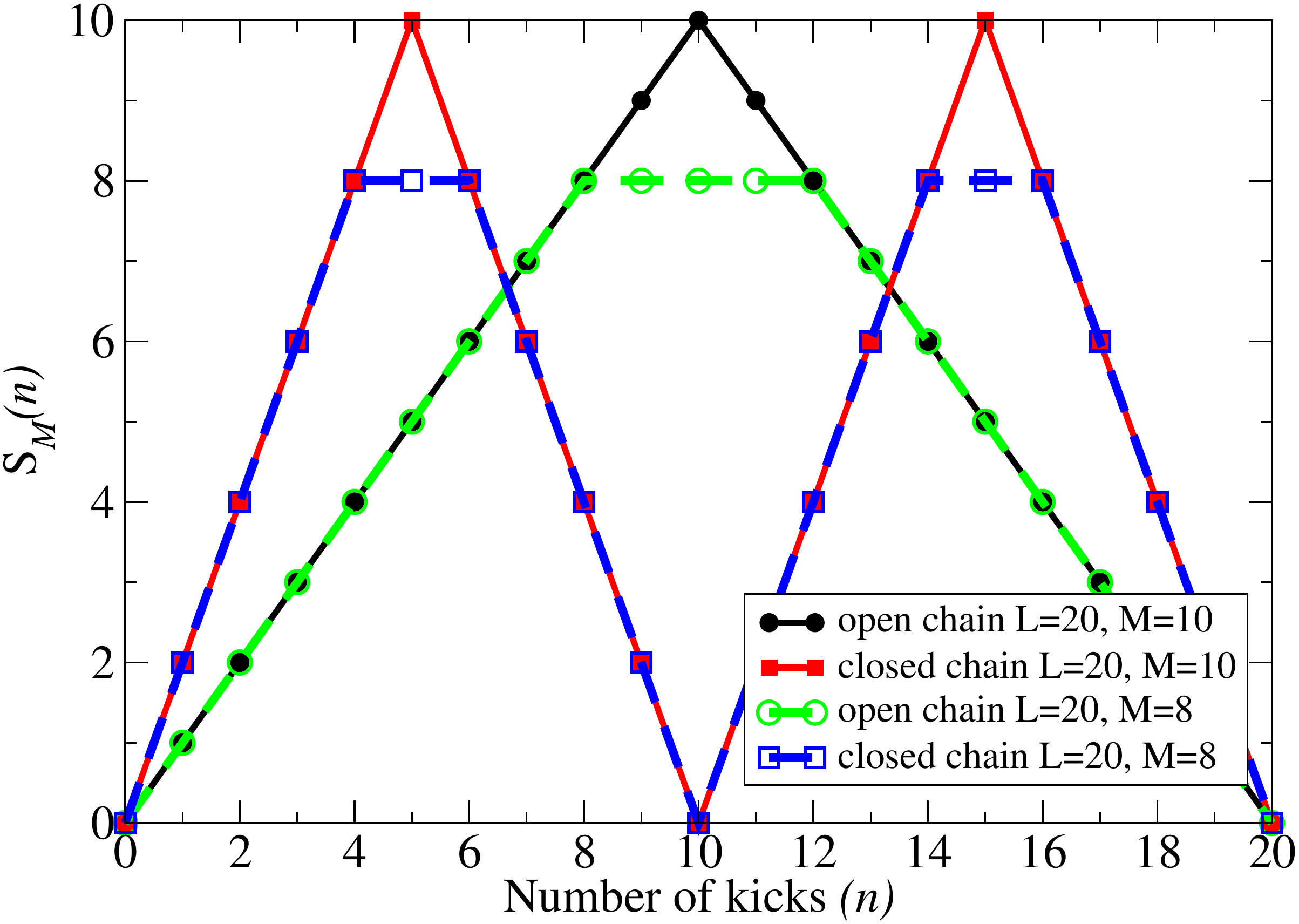}
\caption{Block von Neumann entropy as a function of number of kicks is shown for periodic and open boundary cases for $L=20$. 
The entanglement 
is maximum when $n$ is an odd multiple of $L/2$  ($L/4$)  for open (closed) chain. 
For unequal-size blocks ($M=8, N=12$) the entanglement remains constant 
for $M\le n\le N$ ($M/2\le n\le N/2$) for open (closed) chain.
 }
\label{fig1}
\end{figure}
The block entanglement is a periodic function of $n$, and can be summarized within a period $L$ ($L/2$) for open (closed) chain, with $M=L/2$, as 
\begin{eqnarray}
&S^{\rm Open}_M(n)  = n + (M-n) \Theta(n-M)\Theta{(2M-n)},\\
&S^{\rm Closed}_M(n)  = 2n + 
(M-2n) \Theta(n-{M\over 2})\Theta{(M-n)},
\end{eqnarray}
where $\Theta(x)$ is the Heaviside step function.
The entanglement reaches its maximum possible value $L/2$ at $n=L/2$ in the open chain case, while in the closed this
happens at $n=L/4$ as the Bell pairs are formed from both the interfaces of the blocks, increasing the entanglement 
by 2 ebits every step.  Fig.~\ref{fig1} shows the von Neumann entropy as a function of $n$ in both cases for $L=20$; the 
numerical calculation agrees exactly with the above calculation.

Now, let us consider the case  of unequal  block sizes $M$ (for block A) and $N=L-M$,  let $M<L/2$.
The entanglement increases 1 ebit (2 ebits) for open (closed) chain till $n=M$ ($M/ 2$) as in the case of equal size blocks, and 
it remains constant till $n=N$ ($N/2$), and then decreases by 1 ebit (2 ebits) per step for further kicks.
 The reason for non-decreasing entanglement  at step $n=M+1$ is  due the structure of the operator $V_{M+1}$, given by,
\begin{equation}
V_{M+1}=\exp \left(-i\f{\pi}{4}A^x_{M} B^y_{M+1} B^z_{M}\prod_{j=1}^{M-1}A^z_{j} B^z_{j}\right),
\end{equation}
which is different from the operator shown in Eq.~(\ref{eq:VLby2p1}) (operator when the block sizes are equal). 
This operator will act on the state $|\tilde \psi_{M}\rangle$, with all sites from block $A$ locked in Bell pairs with $L_A$ number of sites
from block $B$, to give the state at the next step $|\tilde \psi_{M+1}\rangle=U_A^{\dg {{M}+1}} U_B^{\dg {{M}+1}}|\psi_{{M}+1}\rangle$, we have,
\begin{eqnarray}
&|\tilde \psi_{M+1}\rangle = |\Phi_{A_1,B_1}..\Phi_{A_{M-1},B_{M-1}}\rangle |0..0\rangle_{B_{M+2}..B_N}\times\nonumber\\
&\left ( |0\rangle_{A_M}|\Phi_{B_M,B_{M+1}}\rangle + |1\rangle_{A_M}|\Phi^{\prime}_{B_M,B_{M+1}}\rangle \right )/ 
\sqrt{2},
\end{eqnarray}
 where we have used another Bell state $|\Phi^\prime\rangle= (|01\rangle - i |10\rangle)/\sqrt{2}$ (See detailed derivation in 
Supplementary Material \cite{suppl}). There are $M-1$ Bell
 pairs between the two blocks, and the spin $A_M$ is entangled with the Bell pair state of the spins $B_M$ and $B_{M+1}$.
It is straightforward to see that the reduced density matrix $\rho_A$ is proportional to the identity matrix, thus the
von Neumann entropy is $M$ at this time step. Similarly, the entropy will stay at this value till $n=N=L-M$, and after that it decreases by 1 ebit (2 ebits) per step for open (closed) chain, showing a clipped sawtooth structure as seen in Fig.~\ref{fig1} for the case of $M=8, L=20$. 

The remarkable algebraic properties of the model considered here lead to the high entanglement content and the creation of Bell pairs up to block-local  operations. However this must {\it not} be construed as the creation of maximum entanglement between qubits, indeed it reflects very well the general observation that high multipartite entanglement, such as between the blocks $A$ and $B$ coexist with low entanglement amongst individual qubits. 
Let us examine how the two-qubit entanglement, viz. concurrence, varies with $n$ by focusing on a pair of qubits $A_i$ and $B_i$, one each from the two blocks. After $n>i$ kicks, but before the unraveling reaches these qubits, the state $|\tilde \psi_n\rangle$
has a Bell pair between the two marked qubits, but the state $|\psi_n\rangle$ has also the nontrivial action of block-local unitaries $U_A$ and $U_B$. 
The reduced density matrix for the marked pair can be written in the form of a particular Krauss representation, 
\begin{equation}
\rho_{_{ A_i B_i}}= \sum_k p_k Q^{A_i}_k(n) Q^{B_i}_k(n) \rho^{\rm Bell}  Q^{A_i \dag}_k(n)  Q^{B_i \dag}_k(n),
\end {equation}
where $Q^{A_i}_k(n)$ are local operators for the qubit $A_i$ 
and $p_k$ are probabilities. These local operators  generates the action of a quantum channel which is in general  decohering and entanglement breaking. 
We have checked numerically that all pair concurrences are zero
for all time steps for both open and  periodic boundary conditions for sizes up to $L=20$. The only exception is the central pair of
qubits $L\over 2$ and ${L\over 2}+1$ in the open chain; this pair has a nonzero concurrence (of unity) at exactly $n={L\over 2}$.
To illustrate the nonzero concurrence for the central pair of qubits, we calculate the concurrence of  the central pair (23) for the 
simple case of $L=4$ explicitly. 
Using Eq.~(\ref{first-step}) and  applying the local unitaries, the state after the first kick is,  
$|\psi_1\kt=\frac{1}{2\sqrt{2}}(-\vert 0 0 0 0\rangle +\vert 0 1 0 1 \rangle +\vert 1 0 1 0\rangle + \vert 1 1 1 1
+ i(\vert 0 0 1 1\rangle+\vert 0 1 1 0\rangle-\vert 1 0 0 1\rangle + \vert 1 1 0 0\rangle  )). $
The reduced density matrix $\rho_{23}$ is proportional to the identity,
which is naturally a separable state. The operators $Q_k$ (with $p_k=1/4$) can be chosen as $Q_1^A=e^{-i\pi \sigma_2^z/4},\,
Q_1^B=e^{-i\pi \sigma_3^z/4}, \, Q_2^A=\sigma_2^x Q_1^A,\,Q_2^B=Q_1^B,\,Q_3^A=Q_1^A,\, Q_3^B=\sigma_3^x Q_1^B, \, Q_4^A=\sigma_2^x Q_1^A,\, Q_4^B=\sigma_3^x Q_1^B$. Apart from a local rotation of the Bell state implemented by $Q_1^A Q_1^B$, the channel is therefore a two-qubit Pauli channel. 
Similarly, the state after the second kick can be written as $|\psi_2\kt=\frac{1}{2}(\vert 0 0 0 0\rangle-i\vert 0 1 1 0\rangle-i\vert 1 0 0 1\rangle-\vert 1 1 1 1\rangle)$, and therefore $\rho_{23}=\frac{1}{2}(\vert 0 0\rangle\langle 0 0 \vert-i\vert 1 1\rangle\langle 0 0 \vert+\vert 1 1\rangle\langle 1 1 \vert
+ i \vert 0 0\rangle\langle 1 1 \vert).$  
For this reduced density matrix, the calculation of concurrence \cite{Wootters1998} gives the maximum value of $1$, as quite simply the state  corresponds to the pure maximally entangled state $(|00\kt -i |11\kt)/\sqrt{2}$. Similarly, the concurrence is zero when $n$ is even multiple of 2, and revives for odd multiples of 2.  
For a periodic chain,  the concurrence remains zero for any number of kicks. 

In conclusion, we have investigated a solvable model that may be interpreted as iteration 
of graph states and shows an entanglement growth of 1 ebit (2 ebits) per iteration for the open (closed) chain,  
 The entanglement generated in the unitary evolution is multipartite in nature.
 On reaching the maximum possible block entanglement, for a contiguous block of $L/2$, it unravels and reaches zero after $L$ iterations.
 These states can be used in protocols of quantum computation, considering their proximity to well investigated graph states. 
 The state after each iteration
 reveals a very interesting algebraic structure through the commuting interaction picture operators $V_i$, that may be hidden
 in approaches using the Jordan-Wigner fermions. This study reveals a further interesting consequence of integrability of the tranverse field Ising model, but we  emphasize that 
the features presented here are unique to the value of the parameter $\tau=\pi/4$, and hence the impulsive field plays an important role, and the 
features are easily revealed while remaining with spin operators rather than fermions.

\pagebreak
\widetext
\begin{center}
\textbf{\large Supplementary Materials: Protocol using kicked Ising dynamics for generating states with maximal multipartite entanglement}
\end{center}
\setcounter{equation}{0}
\setcounter{figure}{0}
\setcounter{table}{0}
\setcounter{page}{1}
\makeatletter
\renewcommand{\theequation}{S\arabic{equation}}
\renewcommand{\thefigure}{S\arabic{figure}}
\renewcommand{\bibnumfmt}[1]{[S#1]}
\renewcommand{\citenumfont}[1]{S#1}

\section{Calculation of $V_n$}
\label{appen_1}
From Eq.~(4) in the main text, we see $V_1=U_B^{\dagger}U_A^{\dagger}X_{AB}U_AU_B$. We have,
 \begin{equation}
 U_A^{\dagger}X_{AB}U_A=Z_A^{\dagger}X_{AA}^{\dagger}\biggl(\frac{1-iA_1^xB_1^x}{\sqrt{2}}\biggr) X_{AA}Z_A.
\end{equation}
As $X_{AA}$ commutes with the central quantity, we have
\begin{eqnarray}
 U_A^{\dagger}X_{AB}U_A=\frac{1}{\sqrt{2}}(1+iA_1^yB_1^x),
\end{eqnarray}
where the identity
$e^{i \f{\pi}{4 }\sigma^z} \sigma^x e^{-i \f{\pi}{4} \sigma^z}=-\sigma^y$
is used. Note the importance of the $\pi/4$ factor.
Similarly the action of  $U_B$ changes $B_1^x$ in $X_{AB}$ to $-B_1^y$, we get
\begin{eqnarray}
 V_1=\biggl(\frac{1-iA_1^yB_1^y}{\sqrt{2}}\biggr)=\exp\left(-i \f{\pi}{4} A_1^yB_1^y\right).
\label{v1eq}
\end{eqnarray}
The second interaction operator
$V_2={U_B^{\dagger}}{U_A^{\dagger}}V_1{U_A}{U_B}$ can be written as
\begin{equation}
 \f{1}{\sqrt{2}}\left(1-i\tilde{A}_1^y\tilde{B}_1^y \right),
\end{equation}
where $\tilde{A}_1^y=Z_A^{\dagger}X_{AA}^{\dagger}A_1^y X_{AA}Z_A$.
Now $X_{AA}^{\dagger}A_1^y X_{AA}=e^{i \f{\pi}{4} A_1^x A_2^x}A_1^y e^{-i \f{\pi}{4} A_1^x A_2^x}=-A_1^z A_2^x$.
which follows from the identity
\beq
e^{i \f{\pi}{4} \sigma_1^x \sigma_2^x } \sigma_1^y  e^{-i \f{\pi}{4} \sigma_1^x \sigma_2^x }=-\sigma_1^z \sigma_2^x.
\eeq
Therefore finally $\tilde{A}_1^y =A_2^y A_1^z$ as 
\beq- e^{i \f{\pi}{4} (\sigma_1^z+\sigma_2^z)} \sigma_1^z \sigma_2^x e^{-i \f{\pi}{4} (\sigma_1^z+\sigma_2^z)}=\sigma_2^y \sigma_1^z.
\eeq
Similarly $B_1^y$ transforms to $B_2^y B_1^z$ and 
 \beq
 V_2=\f{1}{\sqrt{2}}(1-iA_2^yA_1^zB_2^yB_1^z)=\exp\left(-i \f{\pi}{4} A_{2}^yB_2^y A_1^z B_1^z\right).
\eeq

For $2\le n \le L/2-1 $  let 
\beq
V_n =\f{1}{\sqrt{2}}(1-iA_n ^y A_{n-1}^z \cdots A_1^z B_n ^y B_{n-1}^z \cdots B_1^z).
\eeq
From this we show that $U_A^{\dagger} V_n U_A$ has the form of $V_{n+1}$, and hence prove the statement by induction.
Observe that 
\beq
\begin{split}
&U_A^{\dg} A_n ^y A_{n-1}^z \cdots A_1^z U_A=(U_A^{\dg} A_n ^yU_A)(U_A^{\dg} A_{n-1} ^zU_A)\cdots\\ &(U_A^{\dg} A_{2} ^zU_A)(U_A^{\dg} A_{1} ^zU_A)
=(-A_{n+1}^x A_n^x A_{n-1}^y)(A_n^y A_{n-1}^z A_{n-2}^y)\\& (A_{n-1}^y A_{n-2}^z A_{n-3}^y) \cdots  (A_4^y A_3^z A_2^y)(A_3^yA_2^zA_1^y)(-A_2^yA_1^x).
\end{split}
\eeq
The operators in the ``interior" are mapped to a string of three operators, while the ``edges" contribute two. Using properties of Pauli matrices, this simplifies to $(-1)^{n-2} A_{n+1}^y A_n^z A_{n-1}^z \cdots A_1^z$. A similar relation holds for the $B$ operators, and hence finally $V_{n+1}=\f{1}{\sqrt{2}}(1-iA_{n+1} ^y A_{n}^z \cdots A_1^z B_{n+1} ^y B_{n}^z \cdots B_1^z)$, as required.

 The operator $V_{\frac{L}{2}+1}$ for open chain case can also be calculated as above. Observe however that there are now two ``edges" the one with $A_1^z$ and one with $A_{L/2}^y$. One encounters $U_A^{\dg} A_{L/2} ^y A_{L/2-1}^z \cdots A_1^z U_A$ which simplifies to $(-1)^{L/2-2} A_{L/2}^x A_{L/2-1}^z \cdots A_1^z$, with a similar expression for the $B$ string. Thus at this turning point the 
lone $y$ operators turn $x$, and $V_{L/2+1}$ is as given in Eq.~(\ref{eq:VLby2p1}). Further iteration requires $U_A^{\dg} A_{L/2}^x U_A=-A_{L/2}^y$ which along with $U_A^{\dg} A_{L/2-1}^z U_A=A_{L/2}^yA_{L/2-1}^z A_{L/2-2}^y$ results in the decimation of operators at position $L/2$ with the consequence that 
 \beq
 V_{\frac{L}{2}+2}= \exp\left(-i \f{\pi}{4} A^x_{L/2-1} B^x_{L/2-1} \prod_{j=1}^{L/2-2}A^z_{j} B^z_{j} \right).
 \label{eq:VLby2p2}
\eeq
Further decimations lead to $V_L=e^{-i \f{\pi}{4} A_1^x B_1^x}$ and finally $V_{L+1}=V_1=e^{-i \f{\pi}{4} A_1^y B_1^y}$.
\section{Unequal sized blocks}
\label{appen_2}
Let us generalise the formalism in the manuscript by considering unequal size blocks. We divide the chain
into two blocks $A$ and $B$ with $M$ and $N=L-M$ number of spins, respectively, where $M<L/2$.
 Following the discussion in the preceding section for equal sized blocks, the spins in block A $(1,2,\cdots M)$ are relabelled
  as $M, M-1,\cdots 1$ and spins in block B $(M+1,M+2,\cdots M+N)$ are labelled as $1,2,\cdots N$. 
\begin{equation}
\vec  A_j\equiv\vec \sigma_{M+1-j}~~{\rm and}~~
\vec B_j \equiv \vec \sigma_{M+j}.
\label{unequal_AB}
\end{equation}
where the counting of spins is such that $j=1,2,\cdots, M$ in A side and $j=1,2,\cdots, N$ in B side ($N>M$). The 
equal block size can be retraced by $N=M=L/2$. The time evolution of the initial state $|\psi_0\kt=\otimes^L|0\kt$ can be 
calculated by applying powers of the unitary operator given by Eq.~(6) in the main text. Till $n=M$, the time evolved state can be easily
calculated by Eq.~(12) of the main text and the block entanglement is given 
by $S^{\rm open}_M(n)=n$ for open chain case.

Afterwards for $n=M+1$, the state $|\tilde{\psi}_{M+1}\kt$ is calculated by applying 
operator $V_{M+1}=\exp(-i\f{\pi}{4}A^x_{M} B^y_{M+1} B^z_{M}\prod_{j=1}^{M-1}A^z_{j} B^z_{j})$ on $|\tilde{\psi}_{M}\kt$ as
\beq
\begin{split}
|\tilde{\psi}_{M+1}\kt&=\frac{1}{\sqrt{2^{M+1}}}\sum_{\{a_i\}}^{M} \eta(\{a_i\})|a_{M}\cdots 
  a_{1}\rangle_A | a_1  \cdots a_M 0 0 \rangle_B |0\cdots0\rangle_{B_{M+2}..B_N}\nonumber \\
&+(1-2 a_M)|\bar{a}_{M} {a}_{M-1}\cdots 
a_{1}\rangle_A | a_1  \cdots a_M 1 0 \rangle_B |0\cdots0\rangle_{B_{M+2}..B_N}, 
 \end{split}
\eeq     
where the operations $A_M^x|a_M\kt=|\bar{a}_M\kt$ and $B_{M+1}^y|0\kt=i|1\kt$ are performed. Since we are calculating states after
$(M+1)^{\rm th}$ and $(M+2)^{\rm th}$ kicks, therefore, the last two qbits in the block B  $M + 1$ and $M + 2$ are retained and 
rest of the qbits are put together as $|0\cdots0\rangle_{B_{M+2}..B_N}$. Expanding the $M^{\rm th}$ bit in the summation
of above equation leads to 
\beq
\begin{split}
|\tilde{\psi}_{M+1}\kt&=\frac{1}{\sqrt{2^{M+1}}}\sum_{\{a_i\}}^{M-1} \eta(\{a_i\})\bigl(|0 a_{M-1}\cdots 
  a_{1}\rangle_A | a_1  \cdots 0 0 0 \rangle_B +i |1 a_{M-1}\cdots 
  a_{1}\rangle_A | a_1  \cdots 1 0 0 \rangle_B \nonumber \\
&+|1 {a}_{M-1}\cdots a_{1}\rangle_A | a_1  \cdots 0 1 0 \rangle_B-i
|0 {a}_{M-1}\cdots a_{1}\rangle_A | a_1  \cdots 1 1 0 \rangle_B \bigr)|0\cdots0\rangle_{B_{M+2}..B_N},
 \end{split}
\eeq  
which can be rearranged in a nicer form as
\beq
\begin{split}
|\tilde \psi_{M+1}\kt& = |\Phi_{A_1,B_1}\cdots\Phi_{A_{M-1},B_{M-1}}\rangle |0\cdots0\rangle_{B_{M+2}..B_N}\left ( |0\rangle_{A_M}|\Phi_{B_M,B_{M+1}}\rangle + |1\rangle_{A_M}|\Phi^{\prime}_{B_M,B_{M+1}}\rangle \right )/ 
\sqrt{2},
\end{split}
\eeq
where $|\Phi_{l,m}\rangle= (|00\rangle -i |11\rangle)/\sqrt{2}$ and $|\Phi_{l,m}^\prime\rangle= (|01\rangle - i |10\rangle)/\sqrt{2}$ 
are Bell states between a pair of spins located at $l$ and $m$. The reduced density matrix $\rho_A$ (or $\rho_B$) has
$2^M$ equal eigenvalues $1/2^M$ and the block entanglement is $S^{\rm open}_M(M+1)=M$. This shows that $(M+1)^{\rm th}$ kick does not generate
any new ebit between blocks $A$ and $B$ and the entanglement is already exhausted to a saturation value $M$. In order to confirm the saturation of entanglement
 let us calculate one more iteration for $(M+2)^{\rm th}$ kick. The $(M+2)^{\rm th}$ power of unitary operator 
requires $V_{M+2}=U_B^{\dagger}{U_A^{\dagger}}V_{M+1}{U_A}{U_B}$. The calculation of $V_{M+2}$ requires the same method as 
discussed in Section \ref{appen_1} and more specifically needs following operator 
relations $U_A^{\dagger}A_M^xU_A=-A_M^y$, $U_A^{\dagger}A_{M-1}^zU_A=A_M^yA_{M-1}^zA_{M-2}^y$ 
and $U_B^{\dagger} (B^y_{M+1} B^z_{M}\cdots B^z_1) U_B=B^y_{M+2} B^z_{M+1} B^z_{M}\cdots B^z_1$. Using the properties of Pauli matrices, we
can write the expression for $V_{M+2}$ as
\beq
V_{M+2}=\exp(-i\f{\pi}{4}A^x_{M-1} B^y_{M+2} B^z_{M+1}B^z_{M}B^z_{M-1}\prod_{j=1}^{M-2}A^z_{j} B^z_{j}).
\eeq 
Using the relation $|\tilde{\psi}_{M+2}\kt=V_{M+2}V_{M+1}|\tilde{\psi}_{M}\kt$ we calculate the state after $(M+2)^{\rm th}$ kick.
\beq
\begin{split}
 |\tilde{\psi}_{M+2}\kt&=\frac{1}{\sqrt{2^M}}\sum_{\{a_i\}}^{M} \eta(\{a_i\})|a_M a_{M-1}\cdots 
  a_{1}\rangle_A | a_1  \cdots a_M 0 0 \rangle+ |\bar{a}_M \bar{a}_{M-1}a_{M-2}\cdots 
  a_{1}\rangle_A | a_1  \cdots a_M 1 1 \rangle_B\nonumber \\
&+|a_M \bar{a}_{M-1}a_{M-2}\cdots 
  a_{1}\rangle_A | a_1  \cdots a_{M-2} a_{M-1} a_M 0 1 \rangle_B+(1-2a_{M-1}) |\bar{a}_M a_{M-1}a_{M-2}\cdots 
  a_{1}\rangle_A | a_1  \cdots a_M 1 0 \rangle_B. \nonumber \\
\end{split}
\label{psi_m2}
\eeq 
By expanding the $m^{\rm th}$ and $(m-1)^{\rm th}$ bits in the summation, we can express the state as
\beq
\begin{split}
 |\tilde{\psi}_{M+2}\kt&=\frac{1}{\sqrt{2^M}}\sum_{\{a_i\}}^{M} \eta(\{a_i\})|a_{M-2}\cdots 
  a_{1}\rangle_A | a_1  \cdots a_{M-2}\rangle_B  |\chi(a_{M-1},a_M)\kt,
\end{split}
\eeq
or in a more compact manner as
\beq
\begin{split}
|\tilde{\psi}_{M+2}\kt&=|\Phi_{A_1,B_1}\cdots\Phi_{A_{M-2},B_{M-2}}\rangle |\chi(a_{M-1},a_M)\kt|0\cdots0\rangle_{B_{M+2}..B_N},
\end{split}
\eeq
where state $|\chi(a_{M-1},a_M)\kt$ is a composite state of $(M-1)^{\rm th}$ and $M^{\rm th}$ qubits of block A as well as $(M-1)^{\rm th}$, 
$M^{\rm th}$, $(M+1)^{\rm th}$ and $(M+2)^{\rm th}$ qubits of block B and can be easily calculated by expanding Eq.~(\ref{psi_m2}). The explicit 
form of these states are given as  $|\chi(0,0)\kt=\f{1}{2}|0000\kt+|1001\kt+|0110\kt+|1111\kt$,
 $|\chi(0,1)\kt=\f{1}{2}|0001\kt+|1000\kt+|0111\kt-|1110\kt$,  $|\chi(1,0)\kt=\f{1}{2}|0010\kt+|1011\kt+|0100\kt-|1101\kt$ and
 $|\chi(1,1)\kt=\f{1}{2}|0011\kt-|1010\kt+|0101\kt+|1100\kt$. Again, the reduced density matrix has $2^M$ equal eigenvalues $1/2^{M}$ and
block entanglement $S^{\rm open}_M(M+2)=M$. For further kicks the entanglement does not change from the value at kick $n=M$ and remains constant till $n=N$.
Afterwards the entropy starts decreasing in a unit step and reaches to zero after $L$ kicks.  
\end{document}